# ANALYSIS OF THE SEHR SPECTRA OF SYMMETRICAL MOLECULES ON THE BASE OF THE DIPOLE-QUADRUPOLE THEORY

A.M. Polubotko[1*], V.P. Smirnov[2]

[1]A.F. Ioffe Physico-Technical Institute of Russian Academy of Sciences,

Politechnicheskaya 26, 194021 Saint Petersburg, RUSSIA, E-mail: alex.marina@mail.ioffe.ru

,Tel: (812) 274-77-29, Fax: (812) 297-10-17

[2]State University of Information Technologies, Mechanics and Optics, Kronverkskii 49, 197101

Saint Petersburg RUSSIA

**ABSTRACT**

The paper analyses the SEHR spectra of symmetrical molecules on the base of the dipole-quadrupole SEHRS theory. Existence of the bands, caused by vibrations transforming after the unitary irreducible representations of corresponding symmetry groups is demonstrated. As it follows from the theoretical group analysis, these bands are forbidden in phenazine, pyrazine and in usual HRS in molecules with $C_{nh}, D$ and higher symmetry groups. Their appearance strongly confirms the dipole-quadrupole SEHRS theory, which is able to explain other features of the SEHR spectra of these molecules also. Investigation of the SEHR spectra of pyridine and crystal violet demonstrates that they can be explained by this theory too. Thus examination of these spectra strongly confirms the dipole-quadrupole SEHRS theory, which succeeded in explanation of the SERS phenomenon. All these results point out the existence of the strong quadrupole light-molecule interaction, which is the reason of surface-enhanced optical processes.

# Introduction*

In the previous paper[1] we obtained the SEHR cross-section, with the dipole and quadrupole interactions taken into account. It appears that it consists of contributions, which depend on the set of the dipole and quadrupole moments. These contributions follow selection rules

$$\Gamma_{(s,p)} \in \Gamma_{f_1}\Gamma_{f_2}\Gamma_{f_3}, \qquad (1)$$

where the symbols $\Gamma$ designate irreducible representations of ($s, p$) vibrational modes, and of the $f_1, f_2$ and $f_3$ dipole and quadrupole moments. Further we shall consider only the molecules, which belong to the symmetry groups, where all the dipole and quadrupole moments transform after irreducible representations. In accordance with consideration of the dipole and quadrupole interactions[1] it appears that the most enhanced contributions are those, caused by three main quadrupole moments. In addition there is strong enhancement of the contributions, caused by two main quadrupole and one main dipole moment. Besides there will be also several types of contributions, caused by main dipole and quadrupole moments, which are strongly enhanced but with a lesser degree. More detailed information one can find in the previous paper[1]. Here we analyze the specific SEHR spectra of phenazine, pyrazine, trans-1,2-bis (4-pyridyle) ethylene, pyridine and crystal violet. It is demonstrated that appearance of strong bands, caused by totally symmetric vibrations, transforming after the unitary irreducible representations (which are observed in fact in all these molecules) can be explained by this theory, based on the conception of the strong quadrupole light-molecule interaction. In addition it appears that we are able to explain all other details of these spectra. However there are some exclusions, which are associated not with the theory, but with the difficulties of assignment of various Raman bands in large molecules. This notion refers to trans-1, 2-bis (4-pyridyle) ethylene and crystal violet. The other molecules follow the regularities obtained in the previous paper[1] reliably. Let us consider interpretation of the SEHR spectra of these molecules in details.

*The paper is continuation of the previous paper[1]. Therefore here we use brief information, which one should read in[1].

# 1. Analysis of the phenazine SEHR spectrum

The amount of works, which analyze symmetrical properties of the SEHR spectra of symmetrical molecules is very small. The main feature of these works is consideration of the SEHR spectra using the dipole approximation of the light-molecule interaction Hamiltonian only. Let us consider the SEHR spectrum of phenazine[2] which is presented in Fig. 1.

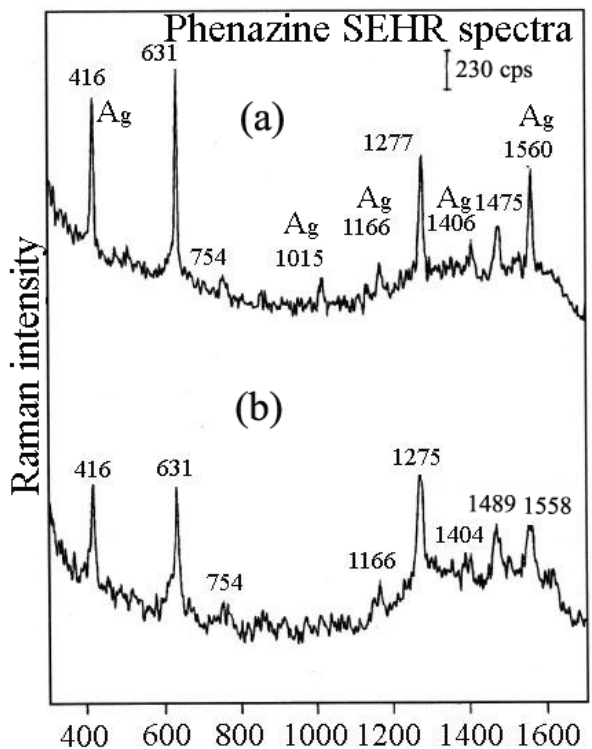

Figure 1. The SEHR spectra of phenazine adsorbed on silver electrode in a solution of saturated phenazine (~ $10^{-4}$ M) + 0.01 M $HClO_4$ + 0.1 M KCl at the potentials (a) 0 V and (b) -0.2 V. Appearance of the bands at 416, 1015, 1166, 1406 and 1560 $cm^{-1}$ with $A_g$ irreducible representation strongly confirms the dipole-quadrupole SEHRS theory.

The main difficulty is assignment of vibrations to various irreducible representations of the $D_{2h}$ symmetry group, which describes symmetry properties of this molecule. Here we use the assignment, which can be found in[3]. However it appears that several calculated modes, which are used for the assignment may belong to various irreducible representations and have very close values[3]. This fact may prevent the assignment of experimental bands unequivocally.

The main feature of the phenazine SEHR spectra both for 0 V and 0.2 V applied potentials[2] is appearance of sufficiently strong bands with $A_g$ symmetry, which are forbidden in usual HRS. They are the bands with 416, 1015, 1166, 1406 and 1560 $cm^{-1}$. Table 1. Here we write out the wave numbers, which refer to the 0 V applied potential.

Table 1. Symmetry and the wavenumbers of observable bands of the SEHR spectrum of phenazine. The values of wavenumbers in parenthesis correspond to those in [3].

| SEHRS | | Assignment | | | |
|---|---|---|---|---|---|
| 0 V $(cm^{-1})$ | -0.2 V $(cm^{-1})$ | $A_g$ | $B_{1u}$ | $B_{2u}$ | $B_{3u}$ |
| 416s | 416s | $A_g$ | | | |
| 631vs | 631vs | | | $B_{2u}$ (657) | |
| 754w | 754vw | | | | $B_{3u}$ (749) |
| 1015w | | $A_g$ | | | |
| 1166w | 1166w | $A_g$ | | | |
| 1277s | 1275s | $A_g$ (1280) | | $B_{2u}$ (1275) | |
| 1406w | 1404w | $A_g$ | | | |
| 1475m | 1469m | $A_g$ (1479) | | $B_{2u}$ (1471) | |
| 1560s | 1558m | $A_g$ | | | |

Enhancement of these bands arises due to the strong dipole and quadrupole light-molecule interactions and is caused mainly by the $(Q_{main} - Q_{main} - Q_{main})$ types of the scatterings for both orientations of the molecules, by $(Q_{main} - d_z - d_z)$ scattering types of horizontally oriented molecules and by $(Q_{main} - d_y - d_y)$ scattering types of vertically adsorbed molecules, when the $d_y$ component of the dipole moment is parallel to the enhanced $E_z$ component of the surface electric field. The bands at 1277 $cm^{-1}$ and 1475 $cm^{-1}$ may refer both to the $A_g$ and $B_{2u}$ symmetry types, because of uncertainty of rigorous assignment of these bands, associated with the close values of the wave numbers[3]. However, they can be explained by $(Q_{main} - Q_{main} - Q_{main})$ and $(Q_{main} - d_z - d_z)$ scattering types for both vertical and horizontal orientations in case of $A_g$ symmetry type and by the $(Q_{main} - Q_{main} - d_y)$ scattering types of horizontally and vertically oriented molecules for the $B_{2u}$ type of symmetry. It should be noted that both types of the lines should be strongly enhanced and they manifest in the SEHR spectrum of this molecule. The band at 754 $cm^{-1}$ of $B_{3u}$ symmetry type can be explained by the $(Q_{main} - Q_{main} - d_x)$ scattering types of horizontally and vertically oriented molecules. This line must be very weak since it is determined by the $E_x$ non-enhanced tangential component of the electric field that is observed in the experiment (Table 1). Thus existence of all the bands of phenazine observed in[2] can be explained by the dipole-quadrupole theory.

## 2. Analysis of the pyrazine SEHR spectrum

Analysis of experimental SEHR spectrum of pyrazine obtained in[4,5] (Fig. 2)

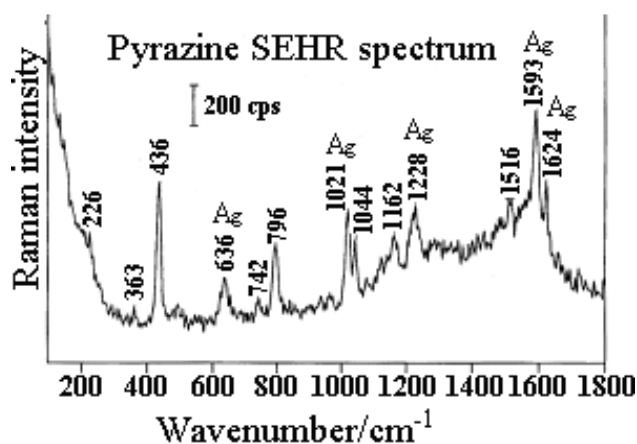

Figure 2. The SEHR spectrum of pyrazine [4]. One can see appearance of the strong bands of $A_g$ symmetry at 636, 1021, 1228, 1593 and 1624 $cm^{-1}$, arising due to the strong quadrupole light-molecule interaction.

also confirms the dipole-quadrupole theory because it allows to explain appearance of strong forbidden bands with $A_g$ symmetry. The SEHR spectra, obtained in[4,5] slightly differ one from another because of some difference in experimental conditions. However, the bands symmetry is well known that is sufficient for the analysis of the SEHR spectra. Below we shall analyze the SEHR spectrum obtained in [4]. Regretfully the authors of [4] did not pointed out the number of adsorbed layers or the coverage of substrate which is necessary for the analysis of real experimental situation. Apparently there were several adsorbed layers. In this case the pyrazine molecules adsorb by three manners because of possible superposition of the molecules: flatly, arbitrary and binding via the nitrogen atom with the surface. Due to these possible orientations all of the $d$ moments can be essential for the scattering. In accordance with the selection rules (1) the lines caused by the vibrations with the following symmetry are observed in the SEHR spectrum Table 2.

Table 2. Symmetry and the wavenumbers of observable bands of the SEHR spectrum of pyrazine.

| Symmetry type | Wavenumbers for SEHRS $(cm^{-1})$ | Relative intensity |
|---|---|---|
| $A_g$ | 1624 | m |
| $A_g$ | 1593 | vs |
| $A_g$ | 1228 | s |
| $A_g$ | 1021 | s |
| $A_g$ | 636 | s |
| $B_{1u}$ | 1044 | m |
| $B_{2u}$ | 796 | s |
| $B_{2u}$ | 436 | vs |
| $B_{3u}$ | 1162 | m |
| $B_{2g}$ | 1516 | w |
| $B_{3g}$ | 742 | w |

1. $A_g$ - (636, 1021, 1228, 1593 and 1624 $cm^{-1}$) caused mainly by $(Q_{main} - Q_{main} - Q_{main})$ and $(Q_{main} - d_z - d_z)$ scattering contributions of flatly adsorbed pyrazine and $(Q_{main} - Q_{main} - Q_{main})$ and $(Q_{main} - d_y - d_y)$ of vertically adsorbed molecules. These lines are forbidden in usual HRS and their appearance strongly proves our point of view.

2. $B_{1u}$ - 1044 $cm^{-1}$ caused mainly by the $(Q_{main} - Q_{main} - d_z)$ and $(d_z - d_z - d_z)$ scattering contributions of horizontally adsorbed pyrazine. Apparently the contributions $(Q_{main} - Q_{main} - d_z), (d_y - d_y - d_z)$ of vertically adsorbed pyrazine are small since they contain the $d_z$ moment, which is associated with the non enhanced $E_y$ tangential component of the electric field for this orientation. Thus the enhancement of the band with $B_{1u}$ symmetry is caused mainly by the above contributions of the horizontally oriented molecules.

3. $B_{2u} - 436$ and 796 $cm^{-1}$ caused mainly by the $(Q_{main} - Q_{main} - d_y)$ and $(d_y - d_y - d_y)$ scattering contributions of vertically adsorbed molecule, when the $d_y$ moment is parallel to the enhanced $E_z$ component of the electric field, which is perpendicular to the surface. The horizontally adsorbed pyrazine apparently does not determine the intensity of the bands of $B_{2u}$ symmetry, since the corresponding contributions $(Q_{main} - Q_{main} - d_y)$ and $(d_z - d_z - d_y)$ include the non enhanced tangential $E_y$ component of the electric field. Thus the enhancement of the bands with $B_{2u}$ symmetry is caused mainly by the above contributions of vertically oriented molecules.

4. $B_{3u}$ - 1162 $cm^{-1}$ caused mainly by the $(Q_{main} - Q_{main} - d_x)$ and $(d_z - d_z - d_x)$ scattering contributions of horizontally adsorbed pyrazine and by $(Q_{main} - Q_{main} - d_x)$ and $(d_y - d_y - d_x)$ scattering contributions of vertically adsorbed pyrazine. Apparently its sufficiently strong enhancement is associated with the superposed molecules oriented arbitrary with the surface, with small deflection from horizontal position. Such deflection apparently impossible for the vertically adsorbed molecules due to sufficiently strong chemical binding of the molecules with the surface.

5. The lines of the $B_{2g}$ and $B_{3g}$ symmetry (1516 and 742 $cm^{-1}$ respectively) may be caused mainly by $(Q_{main} - d_z - d_x)$ and $(Q_{main} - d_z - d_y)$ scattering contributions of the horizontally oriented molecules and are associated with the arbitrary oriented molecules too. Apparently the vertically oriented molecules do not deflect from the vertical position and do not determine intensities of the bands with $B_{2g}$ and $B_{3g}$ symmetry.

The absence of the lines of $A_u$ and $B_{1g}$ symmetry caused by $(d_z - d_x - d_y)$, $(Q_{main} - d_x - d_y)$ and similar scattering contributions may be associated with the small enhancement of these lines in the spectrum due to the non-enhanced components of the electric field, which define the intensity of these contributions.

Thus appearance of all the lines in the SEHR spectrum of pyrazine can be successfully explained by our theory, while appearance of the strong lines with $A_g$ symmetry strongly confirms existence of the strong quadrupole light-molecule interaction in this molecule too.

## 3. Some notions concerning the SEHR spectra of trans-1, 2-bis (4-pyridyle) ethylene and pyridine

It should be noted that there are several works, concerning the analysis of the SEHR spectra of trans-1, 2-bis (4-pyridyle) ethylene[6] and pyridine[4,5,7,8]. The main feature of these works is the analysis of the SEHR spectra using the dipole light-molecule interaction only. In addition the authors of[4,5,6-8] use very approximate numerical methods for calculations of various characteristics of hyper Raman processes like vibrational wavenumbers, polarizabilities, hyperpolarizabilities, their normal coordinate derivatives, band intensities and some others. From our point of view the neglecting by the strong quadrupole light-molecule interaction and the use of these methods result in a very large discrepancy of the experimental results on these spectra and their calculated values. Let us consider the results, which refer to trans-1, 2-bis (4-pyridyle) ethylene in the adsorption geometry, when the long $y$ axis of the molecule is perpendicular to the surface. In accordance with the above SEHRS theory only the bands of $A_u$ and $B_u$ symmetry can be observed in the dipole approximation. This result corresponds to the results and conditions published in[6]. However, consideration of the calculated vibration wavenumbers of the bands with the $A_g$ and $B_u$ symmetry demonstrates nearly the same values for major vibration wavenumbers within the calculation errors (0-10) $cm^{-1}$ (Table 3).

Table 3. Calculated and experimental wavenumbers of the SER and SEHR bands of trans-1, 2-bis (4-pyridyle) ethylene and their symmetry types in accordance with [6].

| The numbers of vibrations of $A_g$ symmetry | Calculated wavenumbers of $A_g$ symmetry $(cm^{-1})$ | The numbers of vibrations of $B_u$ symmetry | Calculated wavenumbers of $B_u$ symmetry $(cm^{-1})$ | Experimental wavenumbers for SEHRS of $B_u$ symmetry $(cm^{-1})$ | Experimental wavenumbers for SERS of $A_g$ symmetry $(cm^{-1})$ |
|---|---|---|---|---|---|
| 17 | 280 | 16 | 466 | 552 | 320 |
| 16 | 645 | 15 | 537 | 599 | 652 |
| 15 | 675 | 14 | 677 | 688 | 663 |
| 14 | 866 | 13 | 814 | 841 | 847 |
| 13 | 986 | 12 | 987 | 972 | 1008 |
| 12 | 1071 | 11 | 1071 | 1007 | 1064 |
| 11 | 1097 | 10 | 1095 | 1116 | |
| 10 | 1145 | 9 | 1137 | 1198 | 1200 |
| 9 | 1201 | 8 | 1221 | 1208 | 1200 |
| 8 | 1224 | 7 | 1239 | 1289 | 1244 |
| 7 | 1340 | 6 | 1302 | 1325 | 1314 |
| 6 | 1363 | 5 | 1366 | 1341 | 1338 |
| 5 | 1409 | 4 | 1418 | 1422 | 1421 |
| 4 | 1498 | 3 | 1504 | 1489 | 1493 |
| 3 | 1560 | 2 | 1569 | 1548 | 1544 |
| 2 | 1609 | 1 | 1610 | 1593 | 1604 |
| 1 | 1682 | | | | 1640 |
| | 3018 | | 3012 | | |
| | 3034 | | 3033 | | |
| | 3049 | | 3049 | | |
| | 3061 | | 3061 | | |
| | 3069 | | 3069 | | |

| The numbers of vibrations of $B_g$ symmetry | Calculated wavenumbers of $B_g$ symmetry ($cm^{-1}$) | The numbers of vibrations of $A_u$ symmetry | Calculated wavenumbers of $A_u$ symmetry ($cm^{-1}$) | Experimental wavenumbers for SEHRS of $A_u$ symmetry ($cm^{-1}$) | Experimental wavenumbers for SERS of $B_g$ symmetry ($cm^{-1}$) |
|---|---|---|---|---|---|
| 8 | 409 | 9 | 298 | 306 | 400 |
| 7 | 503 | 8 | 408 | 385 | 491 |
| 6 | 747 | 7 | 572 | 664 | 738 |
| 5 | 829 | 6 | 761 | 803 | 802 |
| 4 | 900 | 5 | 865 | 878 | 881 |
| 3 | 956 | 4 | 901 |  | 955,972 |
| 2 | 1029 | 3 | 1004 |  | 1064 |
| 1 | 1047 | 2 | 1033 | 1060 |  |
|  |  | 1 | 1047 |  |  |

This notion refers to the wavenumbers with numbers (15-14), (13-12), (12-11), (11-10), (10-9), (8-8), (6-5), (5-4), (4-3), (3-2), (2-1), and to all others in fact. Here the first number refers to the vibrations with $A_g$ and the second with $B_u$ symmetry. In this situation unequivocal assignment of the SEHR bands is impossible since the difference of the calculated wavenumbers with various symmetry is significantly less, than the difference between the measured and calculated wavenumbers. The last one achieves (50-60) $cm^{-1}$. Thus, in accordance with the results of Table 1, a large part of experimental bands, which are assigned to the irreducible representation $B_u$ can be successfully assigned to $A_g$. Moreover, comparison of the wavenumbers of the bands with $B_u$ symmetry measured in SEHRS experiments with the ones of $A_g$ symmetry measured in SERS experiments demonstrates a very small difference ~ (1-10) $cm^{-1}$ for a large number of the bands. Thus both results allow to establish that many SEHRS bands of $B_u$ symmetry can be assigned to $A_g$ symmetry in fact. Taking into account that the bands with $A_g$ symmetry are allowed in the SEHR spectra of 1, 2-bis (4-pyridyle) ethylene in the dipole-quadrupole theory, the above facts do not contradict to explanation of this SEHR spectra in terms of the strong dipole and quadrupole light-molecule interactions.

In order to do this, we must consider real possible orientations of the trans-1, 2-bis (4-pyridyle) ethylene molecules with respect to the surface. In principle, the molecule can adsorb in two main manners-vertically with the main $y$ axis perpendicular to the surface and in flat orientation. Since interaction of the molecule with the surface is weak, one can consider, that the molecule symmetry preserves for both orientations[9,10]. The enhancement of the bands with $A_g$ symmetry is caused mainly by $(Q_{main} - Q_{main} - Q_{main})$, $(Q_{main} - d_z - d_z)$ contributions in the horizontal orientation and by $(Q_{main} - Q_{main} - Q_{main})$ and $(Q_{main} - d_y - d_y)$ in the vertical orientation in accordance with classification of the moments after enhancement degree and the selection rules[1] ((1) in this article). In fact these lines must exist for both orientations. The enhancement of the bands with $A_u$ symmetry is caused mainly by the $(Q_{main} - Q_{main} - d_z)$ and $(d_z - d_z - d_z)$ contributions in the horizontal orientation and by $(Q_{main} - Q_{main} - d_z)$ and $(d_y - d_y - d_z)$ in the vertical orientation, where $d_y$ is perpendicular to the surface and parallel to the $E_z$ component of the electric field. The last contributions are apparently small, because they are caused by $d_z$ moment and by $E_y$ tangential component of the electric field which is not enhanced. The bands with $B_u$ symmetry are caused mainly by $(Q_{main} - Q_{main} - d_{\min or})$ and $(d_z - d_z - d_{\min or})$ scattering types, where under $d_{\min or}$ we mean the $d_x$ and $d_y$ components of the dipole moment in the horizontal orientation, and mainly by $(Q_{main} - Q_{main} - d_y)$ and $(d_y - d_y - d_y)$ contributions, which can be strongly enhanced in the vertical orientation. In principle the bands with $B_g$ symmetry can appear in the SEHRS spectra mainly due to existence of the contributions $(Q_{main} - d_z - d_x)$ and $(Q_{main} - d_z - d_y)$ in the horizontal orientation and mainly due to $(Q_{main} - d_y - d_z)$ contributions in the vertical orientation. Analysis of experimental SERS and SEHRS bands demonstrates existence of very close values of wavenumbers of the bands with $A_g$ and $B_g$ symmetry in SERS and of $B_u$ and $A_u$ symmetry in SEHRS (Table 1). Taking into account the large uncertainty in determination of symmetry of the SERS and SEHRS bands due to the calculation errors, one can make a conclusion, that a part of the SEHRS lines can belong to $A_g$ and $B_g$ symmetry. Regretfully we are not able to explain the SEHRS results of [6] with better accuracy because of crude approximation made in calculations of the wavenumbers and in determination of the bands symmetry in this work. However, consideration of the strong quadrupole light-molecule interaction and of the two possible orientations of the molecules allows to explain a possibility of appearance of the bands with $A_g$ and $B_g$ symmetry in the SEHR spectra of trans-1, 2-bis (4-pyridyle) ethylene.

The SEHR spectra of pyridine were presented in[4,5,7,8]. In principle, the spectra differ slightly one from another that can be explained by various experimental conditions. However, the positions of the observed SEHRS lines are nearly the same for all these spectra. Therefore we will analyze the spectrum, obtained in [7] (Fig. 3).

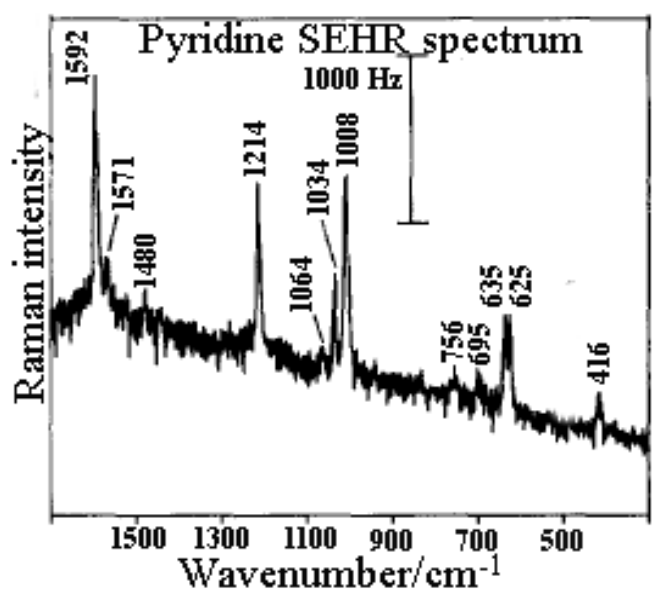

Figure 3. The SEHR spectrum of pyridine [7].

The dipole approximation is able to explain main peculiarities of the SEHR spectrum of pyridine. This fact is associated with specific symmetry of this molecule and with the fact that the $d_z$ moment of the molecule transforms in accordance with the unitary irreducible representation of the $C_{2v}$ symmetry group. It is the reason that all the observed lines are allowed in the dipole approximation. Therefore, pyridine is not the molecule which succeeds in discovery of the strong quadrupole light-molecule interaction. However, we can reconsider the SEHRS results of [7,8] on the base of the dipole-quadrupole theory and explain all the features of the SEHR spectra. Firstly, one should note that the calculated wavenumbers of the pyridine vibrations, obtained in [7] differ strongly from the values, observed experimentally from the SEHR spectrum. This result is apparently associated with the use of very approximate methods of determination of vibrational numbers in[7]. However, the experimental wavenumbers well fit the values, published in [11] for the SERS spectrum of pyridine on $Ag$ with 0.2L exposure. Therefore, we use the values, published in[11] for the symmetry analysis. The pyridine molecule adsorbs primary vertically, binding with the surface via nitrogen. However, a part of the molecules can adsorb horizontally, or in arbitrary manner due to possible superposition of these molecules in the first layer. Then in accordance with our theory all the $Q_1 = Q_{xx}, Q_2 = Q_{yy}, Q_3 = Q_{zz}$ and $d_x, d_y, d_z$ moments may be essential for the scattering and can be the main moments for this system for various configurations. However, the influence of the $d_x$ moment is apparently significantly less than of $d_y$ and $d_z$, since superposition of two pyridine molecules does not change flat orientation of the molecule strongly. Since $Q_1, Q_2, Q_3$ and $d_z$ moments transform after the unitary irreducible representation, in accordance with the selection rules[1] ((1) in this article), the most enhanced contributions $(Q_{main} - Q_{main} - Q_{main}), (Q_{main} - Q_{main} - d_z), (Q_{main} - d_z - d_z)$ and $(d_z - d_z - d_z)$ define strongest enhancement of the bands, caused by the totally symmetric vibrations with $A_1$ symmetry (Table 4). Appearance of the bands caused by vibrations with $B_2$ symmetry is caused mainly by the $(Q_{main} - Q_{main} - d_y)$ type of the scattering for pyridine lying flatly at the surface.

Table 4. Symmetry and wavenumbers of observable bands of the SEHR spectrum of pyridine.

| Symmetry | The mode number | Wavenumbers ($cm^{-1}$) [7] | Relative Intensity |
|---|---|---|---|
| $A_1$ | 3 | 625 | s |
| | 1 | 1008 | vs |
| | 6 | 1034 | s |
| | 8 | 1064 | vw |
| | 5 | 1214 | vs |
| | 9 | 1480 | vw |
| | 4 | 1592 | vs |
| $B_1$ | 14 | 1571 | wv |
| $B_2$ | 27 | 416 | w |
| | 26 | 695 | w |
| | 23 | 756 | w |

In this case the $d_y$ moment, which is perpendicular to the pyridine plane is parallel to the enhanced $E_z$ component of the local electric field and the overall scattering becomes sufficiently strong. The appearance of the bands, caused by the vibrations with $B_1$ symmetry, is apparently caused by $(Q_{main} - Q_{main} - d_x)$ scattering from molecules superposed one on another, which lie nearly flatly on the surface. Since the primary orientation of the adsorbed pyridine is vertical, the number of other orientations apparently is small and the intensity of the bands with $B_1$ and $B_2$ symmetry is small too. One should note that there is a sufficiently strong line at $635 cm^{-1}$ in the experimental SEHR spectra of pyridine, published in[7,8]. This line is absent in the pyridine spectra in[4,5] and among the calculated wave-numbers for pyridine in[7,8]. Therefore, we do not discuss the origin of this line in the paper. Thus, after the analysis of the SEHR spectrum of pyridine we can assert that the dipole-quadrupole theory successfully explains its SEHR spectrum.

## 4. Some notions about SEHR spectra of crystal violet

One of symmetrical molecules, where the SEHR and SER spectra were investigated is crystal violet (CV). The most reliable assignment of vibrations was made in[12] where the molecule is referred to the $D_3$ symmetry group.

(However one should note, that the authors point out the existence of $A$ irreducible representation, while there is no such representation in this group. There exist $A_1$ and $A_2$ irreducible representations. Apparently the authors had in mind the $A_1$ irreducible representation.)

In[13] IR, SER and SEHR spectra were partially investigated, however the authors considered that the molecule refers to the $C_3$ symmetry group. Recently the authors of[14,15] investigated SER, SEHR, usual Raman, IR, and hyper Raman spectra, considering that the molecule possesses by $D_3$ symmetry, such as in[12]. In addition they made their own assignment on the base of the DFT and B3LYP-311G methods. One should note that careful consideration of experimental and calculated vibrational bands, vibrational wavenumbers, and possible assignment published in all these papers, reveals large discrepancy in calculated values and assignment of the bands (see[12-15]). Therefore, our attempts to make our own assignment of the SEHR bands failed in fact and we consider that the matter of the precise assignment is spurious now. However, there are several bands, which assigned to the $A_1$ irreducible representation in all these works. They are the bands at 1621 and 1389 $cm^{-1}$. In addition comparison of the IR, SER and SEHR spectra, published in[13] points out that the bands 1298, 1370, 1478 and 1588 $cm^{-1}$ are observed in all these spectra. Therefore, in accordance with selection rules in IR spectra these bands can be caused only by vibrations with $A_2$ and $E$ symmetry. Consideration of the SER and SEHR spectra in accordance with the dipole-quadrupole theory[9] demonstrates that these bands may be caused preferably by the $(Q_{main} - d_z)$ and $(Q_{main} - Q_{main} - d_z)$ contributions in SERS and SEHRS respectively. This refers first of all to the bands at 1370 and 1589 $cm^{-1}$ because of their large relative intensity in the SER and SEHR spectra. The assignment of these bands to the $E$ symmetry may also be possible since the bands may be caused by the $(Q_{main} - (d_x, d_y))$ and $(Q_{main} - Q_{main} - (d_x, d_y))$ contributions, for molecules, superposed in the first and second layers having orientation which is not parallel to the surface. As it was considered earlier, this fact depends on the coverage of substrate which is not known in[14]. Thus apparently there is a multilayer coverage in these experiments. Appearance and existence of the bands with $A_1$ symmetry in the SEHR spectra[13-15] is in agreement with our theory. One should note that appearance of the bands in the $D_3$ symmetry group, caused by vibrations with all possible irreducible representations in SEHRS formally is possible in a pure dipole theory, such as in pyridine. However, appearance of the bands with $A_2$ symmetry in SERS can be explained only by the $(Q_{main} - d_z)$ type of scattering. Therefore, the correct explanation of the SEHR spectra of CV should be made on the base of the dipole-quadrupole theory, which apparently is able to explain all other features of the SER and SEHR spectra of CV also. One should note that the SEHR spectra of crystal violet in the above mentioned works sometimes were obtained with the incident radiation, when the resonance scattering is possible. It can be shown that the resonance character of these processes must not change the selection rules[1] ((1) in this paper) for the contributions. Therefore, the result that the bands caused by vibrations with $A_1$ and $A_2$ symmetry can be observed should be valid in case of surface enhanced resonance hyper Raman scattering too, that strongly support our theory.